\documentclass[aps,prc,twocolumn,superscriptaddress,showpacs,floatfix]{revtex4-1}

\usepackage[T1]{fontenc}
\usepackage{latexsym}
\usepackage{graphicx,amssymb}
\usepackage[fleqn]{amsmath}
\usepackage{amsfonts}
\usepackage{dcolumn}
\usepackage{bm}
\usepackage{braket}
\usepackage{multirow}
\usepackage{MnSymbol}
\usepackage{color}
\usepackage{ulem}
\usepackage{hyperref}
\usepackage{bm}

\newcommand{\intt}
{
	\displaystyle \int
}

\begin{document}

\preprint{ }

\title{Gamow shell model description of the radiative capture reaction $^8$B$(p,\gamma)$$^9$C }

\author{G.X. Dong}
\affiliation{School of Science, Huzhou University, Huzhou 313000, China}

\author{X.B. Wang}
\affiliation{School of Science, Huzhou University, Huzhou 313000, China}

\author{N. Michel}
\affiliation{Institute of Modern Physics, Chinese Academy of Sciences, Lanzhou, Gansu 730000, China}

\author{M. P{\l}oszajczak}
\thanks{ploszajczak@ganil.fr}
\affiliation{Grand Acc\'el\'erateur National d'Ions Lourds (GANIL), CEA/DSM - CNRS/IN2P3,
BP 55027, F-14076 Caen Cedex, France}

\date{\today}

\begin{abstract}
\noindent
  {{\bf Background:} In low metallicity supermassive stars, the hot $pp$ chain can serve as an alternative to produce the CNO nuclei. In the astrophysical environment of high temperature, the proton capture of $^8$B can be faster than its beta decay, so that the $^8$B$(p,\gamma)$$^9$C reaction plays an important role in the hot $pp$ chain. Due to the unstable nature of $^8$B and the unavailability of $^8$B beams, the study of the $^8$B$(p,\gamma)$$^9$C reaction can only be achieved by indirect methods, so that large uncertainties exist. \\
   {\bf Purpose:} The Gamow shell model in the coupled-channel representation (GSM-CC) is applied to study the proton radiative capture reaction $^8$B$(p,\gamma)$$^9$C.  \\
  {\bf Method:} The GSM-CC is a unified microscopic theory for the description of nuclear structure and nuclear reaction properties. A translationally invariant Hamiltonian is considered for that matter, making use of a finite-range two-body interaction, whose parameters are adjusted to reproduce the low-energy spectra of $^{8}$B and $^{9}$C. The reaction channels are then built through the coupling of the wave functions of the ground state $2_1^+$, the first excited state $1^+_1$, and the second excited state $3^+_1$ in $^8$B to a projectile proton wave function in different partial waves. For the calculation of the $^8$B$(p,\gamma)$$^9$C astrophysical factor, all E1, M1, and E2 transitions from the initial continuum states to the final bound states ${3/2}_1^-$of $^9$C are considered. The resonant capture to the first resonant state ${1/2}_1^-$ of $^{9}$C is also calculated.  \\
  {\bf Results:} The experimental low-energy levels and the proton emission threshold in $^9$C are reproduced by the GSM-CC. The calculated astrophysical factor agrees with  experimental data obtained from indirect measurements. The reaction rates from the direct capture and resonant capture are calculated for the temperature range of astrophysical interest. \\
  {\bf Conclusion:} The calculated total astrophysical $S$ factor is dominated by the E1 transition to the ground state of $^9$C. The GSM-CC calculations suggest that $S$ first increases with the energy of the center of mass $E_{\rm c.m.}$, and then decrease with the energy. This agrees with existing data, which has smaller values around zero energy and larger values in the energy range of 0.2 MeV $\leq E_{\rm c.m.} \leq$ 0.6 MeV.}

\end{abstract}



\pacs{03.65.Nk, 
	31.15.-p, 
	31.15.V-, 
	33.15.Ry 
}

\maketitle
\section{Introduction}
\label{intro}
The nucleosynthesis of light nuclei is hindered by the mass gap of A=8 nuclear systems, where no stable nucleus exists. However, this gap can be overcome in some astrophysical situations. It is well known that the radiative proton capture reaction $^8$B$(p,\gamma)$$^9$C plays an important role in the hot proton-proton chain in low-metallicity supermassive stars~\cite{Wiescher89,Fuller86}. If density is sufficiently high ($\geq 2\times 10^7$ g/cm$^3$) and temperature is in the range $0.07 \leq {\rm T}_9 \leq 0.7$, proton capture and alpha capture can be faster than beta decay~\cite{Wiescher89}. Thus, this reaction can serve as an alternative to the three-alpha process to form CNO nuclei, through the following reaction chain \cite{Wiescher89}: $^7$Be$(p,\gamma)$$^8$B$(p,\gamma)$$^9$C($\alpha,p$)$^{12}$N$(p,\gamma)$$^{13}$O($\beta^+$)$^{13}$N$(p,\gamma)$$^{14}$O.

Charged-particle capture cross sections are hindered by Coulomb repulsion. Because of the very small reaction cross section, and low intensity of the $^8$B beam, the direct measurement of $^8$B$(p,\gamma)$$^9$C reaction is difficult if not impossible. There are usually two different ways of indirect measurements of this reaction. One makes use of the asymptotic normalization coefficients (ANC) for virtual decay, from which the astrophysical $S$ factor can be deduced. The other is Coulomb dissociation induced by radioactive beams.

There are several experimental studies dealing with the $^8$B$(p,\gamma)$$^9$C reaction based on the ANC method. Beaumel et al. measured the $^8$B$(d,n)$$^9$C reaction cross section with inverse kinematics at 14.4 MeV/u and derived the ANC of the virtual decay of $^9$C$\rightarrow$$^8$B+$p$ using distorted-wave Born approximation (DWBA)~\cite{Beaumel01}. A similar analysis has been performed later at center of mass (c.m.) energy $E_{\rm c.m.}$=7.8 MeV by Guo et al.~\cite{BGuo05}.

Trache et al.~investigated the breakup of $^9$C at intermediate energies to extract the ANC and could estimate the astrophysical $S$ factor using the Glauber model~\cite{Trache02}. Recently, this experiment was re-analyzed using the continuum discretized coupled channel (CDCC) approach and the eikonal reaction theory (ERT)~\cite{Fukui12}. The value of $S$(0) obtained in this analysis was found larger by about 43$\%$ than previously reported~\cite{Trache02}.

Based on the CDCC approach, Fukui et al.~re-analyzed the $^8$B$(d,n)$$^9$C measurement~\cite{Beaumel01}, and obtained a smaller value of the $S$ factor than that of the DWBA analysis of Ref.\cite{Fukui15}. In the CDCC frame, they also analyzed the three-body features of $^9$C in breakup reaction~\cite{Singh}.

Finally, Motobayashi determined the $S$ factor of the $^8$B$(p,\gamma)$$^9$C reaction in a Coulomb dissociation experiment. A $^9$C projectile beam is used therein along with a lead target to generate Coulomb excitation, which leads to an unbound state decaying in the $^8$B+$p$ channel~\cite{Moto03}. The $S$ factor in the energy range 0.2 MeV $\leq E_{\rm c.m.} \leq$ 0.6 MeV reported in this analysis is significantly larger than that found in other experimental analyses at $E_{\rm c.m.}$=0~\cite{Enders03,Trache02,Beaumel01,BGuo05,Fukui12,Fukui15}.

Theoretical calculations of the $S$ factor in the $^8$B$(p,\gamma)$$^9$C reaction exhibit large discrepancies. A very large value of $S(0)$ ($\sim$210 eV barn) was obtained in a simple potential model~\cite{Wiescher89}. Studies of the
$^8$B$(p,\gamma)$$^9$C reaction using the microscopic cluster model have shown a significant dependence of the $S$ factor on the considered potential for that matter ~\cite{Descouvemont93,Descouvemont99}.
Mirror reactions $^8$Li$(n,\gamma)$$^9$Li and $^8$B$(p,\gamma)$$^9$C were studied by P.~Mohr using the direct capture model~\cite{Mohr03}. It was found that cross sections strongly depend on the parameters of the potential, which could be explained by the resonant states present in the potential model.

In this paper, we focus on the study of the proton radiative capture reaction $^8$B$(p,\gamma)$$^9$C in the general framework of the Gamow shell model (GSM)~\cite{Michel02,Michel03,Michel09,Jaganathen14,Fossez15,GSMbook}.
In this model, the many-body states are described as a linear combination of Slater determinants built by the single-particle (s.p.) bound, resonance and non-resonant scattering states of the Berggren ensemble~\cite{rf:4}.
The GSM is an excellent tool for nuclear structure studies~\cite{GSMbook}. However, there is no separation between different reactions channels in the Slater determinant representation. Hence, the direct application of the GSM to study nuclear reactions is not possible. This problem can be solved by formulating GSM in the coupled-channel representation (GSM-CC), which provides a unified theory of nuclear structure and reactions~\cite{Jaganathen14,Betan,Wang21,GSMbook}. The GSM-CC was applied to several reaction studies, e.g., low-energy elastic and inelastic proton scattering~\cite{Jaganathen14}, deuteron elastic scattering~\cite{Mercenne19}, and radiative capture reactions~\cite{Fossez15,dong17,dong22}.

This paper is organized as follows. The formalism of GSM-CC is briefly introduced in Sec.~\ref{sec-2}. Results of GSM-CC calculations are presented and discussed in Sec.~\ref{sec3}. The low-energy spectra of $^8$B and $^{9}$C are depicted in Sec.~\ref{sec3-a} and the low-energy astrophysical factor of the $^8$B$(p,\gamma)$$^9$C reaction is studied in Sec. \ref{sec3-b}. The astrophysical reaction rates are discussed in Sec.~\ref{sec3-c}. Finally, summary and main conclusions are provided in Sec.~\ref{sec4}.

\section{The Gamow shell model in the coupled-channel representation}
\label{sec-2}

In this section, we briefly introduce the GSM-CC approach. For a full presentation of the formalism, see Refs.~\cite{GSMbook,Michel09} and references cited therein.

In GSM and GSM-CC, the Hamiltonian is defined with the intrinsic nucleon-core coordinates of the cluster-orbital shell model (COSM)~\cite{Ikeda88}:
\begin{equation}
	\hat{H} = \sum_{i = 1}^{ {N}_{ \text{val} } } \left( \frac{ \hat{\vec{p}}_{i}^{2} }{ 2 { \mu }_{i} } + {U}_{\text{core}} ( \hat{r}_{i} ) \right) + \sum_{i < j}^{ {N}_{ \text{val} } } \left( V _{ij} + \frac{ {\hat{\vec{p}}_{i}}{\cdot} {\hat{\vec{p}}_{j} }}{ {M}_{\text{core}} } \right)~,
	\label{GSM_Hamiltonian}
\end{equation}
in which $N_\text{val}$ stands for valence particle number, $M_{\text{core}}$ is the mass of the core, $\mu_i$ indicates the reduced mass of the proton or neutron, and $U_{\text{core}}(\hat{r})$ denotes the s.p.~potential of the core acting on valence nucleons. $V_{ij}$ is the two-body interaction between valence nucleons and is translationally invariant. In the above formula, the last term is the recoil term~\cite{GSMbook}. The fundamental advantage of COSM is that it allows to remove spurious c.m.~excitations in many-body wave functions~\cite{Ikeda88}:

In GSM-CC, the ${ A }$-body system is decomposed in reaction channels:
\begin{equation}
  \ket{ { \Psi }_{ M }^{ J } } = \sum_{ {c} } \int_{ 0 }^{ +\infty } \ket{{ \left( r,c \right) }_{ M }^{ J } } \frac{ { u }_{ {c} }^{JM} (r) }{ r } { r }^{ 2 } ~ dr \; ,
  \label{scat_A_body_compound}
\end{equation}
where ${ {u}_{ {c} }^{JM}(r) }$ is the radial amplitude describing the relative motion between the target core and projectile in the channel $c$. It is obtained by solving the GSM coupled-channel equations for fixed total angular momentum $ {J} $ and projection $ {M} $. The variable ${ r }$ in the above equation denotes the relative distance between the c.m.~of the target core and the projectile. $\ket{ \left( r,c \right)^J_M} $ is the binary-cluster channel state defined as:
\begin{equation}
  \ket{ \left( r,c \right)^J_M}  = \hat{ \mathcal{A}} \ket{ \{ \ket{ \Psi_{ {\rm T} }^{J_{ {\rm T} } }  }
  \otimes \ket{ {\Psi^{J_{{\rm p}}}_{{\rm p}}}} \}_{ M }^{J} } \; ,
  \label{channel}
\end{equation}
where channel index $c$ indicates both mass partitions and quantum numbers and the $r$ variable is implicitly considered in $\ket{ {\Psi_{{\rm p}}^{J_{{\rm p}}}}} $. ${\hat{ \mathcal{A}}}$ is the antisymmetrizer acting on nucleons belonging to different clusters. $\ket{\Psi_{ {\rm T} }^{J_{ {\rm T} }} }$ and $\ket{\Psi_{ {\rm p} }^{J_{ {\rm p} }} }$ are the target and projectile states with their angular momentum ${ { J }_{ {\rm T} } }$ and ${ { J }_{ {\rm p} } }$, respectively. The coupling of ${ { J }_{ {\rm T} } }$ and ${ { J }_{ {\rm p} } }$ provides the total angular momentum ${  J }$. $JM$-dependence of channel $c$ will now be omitted for convenience.

The Schr{\"o}dinger equation represented with coupled-channel equations reads:
\begin{equation}
	\sumint\limits_{c} \intt_{0}^{ \infty } dr \, {r}^{2} \left( {H}_{ c' , c } ( r' , r ) - E {O}_{ c' , c } ( r' , r ) \right) \frac{ {u}_{c} (r) }{r} = 0 \; ,
	\label{eq_CC_eqs_general}
\end{equation}
where $E$ is the energy of the A-body system, and the kernels are the Hamiltonian and the norm matrix elements in the channel representation, respectively:
\begin{equation}
	{H}_{ c' , c } ( r' , r ) = \braket{ r' , c' | \hat{H} | r , c } \; ,
	\label{eq_CC_H_ME_general}
\end{equation}
and
\begin{equation}
	{O}_{ c' , c } ( r' , r ) = \braket{ r' , c' | r , c } \; .
	\label{eq_CC_O_ME_general}
\end{equation}

The channel state $\ket{ r , c }$ can be built from a complete Berggren s.p.~basis~\cite{rf:4} containing the bound, resonance, and non-resonant scattering states from the contour in the complex $k$-plane~\cite{Michel02,Michel03,Michel09,GSMbook}:
\begin{equation}
	\ket{ r , c } = \sum_{i} \frac{ {u}_{i} (r) }{r} \ket{ { \phi }_{i}^{ \text{rad} } , c } \; , \label{eq_CC_basis_state_expansion_Berggren_2}
\end{equation}
in which ${ \ket{ { \phi }_{i}^{ \text{rad} } , c } = \hat{ \mathcal{A} } ( \ket{ { \phi }_{i}^{ \text{rad} } } \otimes \ket{c} ) }$ and $\braket{ r| \phi_i^{ \text{rad}}} = {u}_{i} (r) / r $ is the radial part.

The channel basis states $\ket{ r , c }$ are non-orthogonal, which results from the antisymmetrization between projectile and target states. Orthogonalized channel basis states are obtained through the definition:
\begin{equation}
\ket{ r , c }_{o} = \hat{O}^{ -\frac{1}{2} } \ket{ r , c } \; ,
\label{eq_CC_non_ortho_to_ortho_channel_states}
\end{equation}
where ${ \hat{O} }$ is the overlap operator.
Consequently, orthogonalized channels verify:
\begin{equation}
	{}_{\rm o} \braket{ r' , c' | r , c }_{\rm o} = \frac{ \delta ( r' - r ) }{r'r} \delta_{ c' c } \label{eq_CC_ortho_channel_basis_braket} \ .
\end{equation}
The GSM-CC equations~\eqref{eq_CC_eqs_general} thus write:
\begin{align}
	\sumint\limits_{c} \intt_{0}^{ \infty } dr \, {r}^{2} &( {}_{\rm o}\braket{ r' , c' | \hat{H}_{\rm o} | r , c }_{\rm o} - E {}_{\rm o}\braket{ r' , c' | \hat{O} | r , c }_{\rm o} ) \nonumber \\
	&\times {}_{\rm o}\braket{ r , c | { \Psi }_{\rm o} } = 0 \; ,
	\label{eq_CC_eqs_general_clear_ortho}
\end{align}
in which
${ \hat{H}_{\rm o} = \hat{O}^{  \frac{1}{2} } \hat{H} \hat{O}^{  \frac{1}{2} } }$, and $ {\ket{ \Psi_{\rm o} } = \hat{O}^{1/2} \ket{ \Psi } }$.

Assuming ${ \ket{ \Phi } = \hat{O} \ket{ \Psi } }$, the generalized eigenvalue problem given in Eq.~\eqref{eq_CC_eqs_general_clear_ortho} can be put into standard form:
\begin{align}
	\sumint\limits_{c} \intt_{0}^{ \infty } dr \, {r}^{2} & ( {}_{{\rm o}}\braket{ r' , c' | \hat{H} | r , c }_{\rm o} - E {}_{\rm o}\braket{ r' , c' | r , c }_{\rm o} ) \nonumber \\
    &\times {}_{\rm o}\braket{ r , c | \Phi } = 0 \ .
	\label{eq_CC_eqs_general_clear_ortho_again}
\end{align}
One can then write Eq.~\eqref{eq_CC_eqs_general_clear_ortho_again} using channel radial wave functions associated to orthogonalized channel basis states:
\begin{equation}
	\sumint\limits_{c} \intt_{0}^{ \infty } dr \, {r}^{2} \braket{ r' , c' | \hat{H}_{\rm m} | r , c } \frac{ {w}_{c} (r) }{r} = E \frac{ {w}_{ c' } ( r' ) }{ r' }\ ,
	\label{eq_CC_final}
\end{equation}
where ${ \hat{H}_{\rm m} = \hat{O}^{ - \frac{1}{2} } \hat{H} \hat{O}^{ - \frac{1}{2} } }$ and $${w}_{c} (r)/ r \equiv \braket{ r , c | \hat{O}^{ \frac{1}{2} } | \Psi } = {}_{\rm o}\braket{ r , c | \Phi } \ .$$
Because of the completeness of the Berggren basis, these coupled-channel equations (\ref{eq_CC_final}) can be solved numerically from the Berggren basis expansion of the Green's function ${ { (H - E) }^{ -1 } }$ (see Refs.\cite{Mercenne19,GSMbook} for details).

\section{Discussion and results}
\label{sec3}

In GSM and GSM-CC calculations, we assume $^4$He as the inert core. $^8$B and $^9$C will then consist of 3 and 4 weakly bound or unbound valence protons above the  $^4$He core, respectively, along with one well-bound valence neutron. The core potential in the Hamiltonian is given by the Woods-Saxon (WS) potential with a spin-orbit term:
\begin{equation}
  V_{WS}(r) = V_o f(r) - 4V_{so} \frac{1}{r} \frac{df(r)}{dr}\, \vec{\ell} \cdot \vec{s} + V_{\rm Coul}(r) \ ,
  \label{eq.woodsSaxon}
\end{equation}
where
\begin{equation}
f(r) = \frac{1}{1 + \exp((r-R_0)/a)} \label{WS_form_factor}
\end{equation}
In these equations, $\vec{\ell}$ and $\vec{s}$ are the orbital and spin angular momenta, respectively, $a$ is the diffuseness of the WS potential, $R_0$ is its radius,
$V_o$ and $V_{so}$ are its central and spin-orbit strength, respectively. The Coulomb potential for protons $V_{\rm Coul}(r)$ is generated by a spherical Gaussian charge distribution of radius $R_{ch}$ and reads: \begin{equation}
V_{\rm Coul}(r) = 2e^2\: \frac{1}{r}\mbox{erf}(r/\tilde{R}_{ch})  \ , \label{Coulomb_potential}
\end{equation}
where one has introduced the modified charge radius $\tilde{R}_{ch} = 4R_{ch}/(3\sqrt{\pi})$, which allows for $V_{\rm Coul}(r)$ to close resemble the Coulomb potential generated by a uniformly-charged sphere of radius $R_{ch}$. The numerical values of the used parameters in the WS potential can be found in Table~\ref{para-1}.
\begin{table}[ht!]
\caption{Parameters of the WS potential of the $^4$He core used in the description of $^8$B, $^9$C spectra, and proton-capture cross-section $^{8}\text{B}(p,\gamma)^{9}\text{C}$. Only the central potential depth $V_\text{o}$ is $\ell$-dependent. See Eqs.~\ref{eq.woodsSaxon}-\ref{Coulomb_potential} for notations. \label{para-1}}
\begin{ruledtabular}
\begin{tabular}{ccc}
  Parameter & Protons & Neutrons \\
  \hline
  $a$ & 0.65 fm & 0.65 fm \\
  $R_0$ & 2 fm & 2 fm \\
  $R_{ch}$ & 2.54 fm & --- \\
  $V_\text{so}$ & 8.5 MeV & 8.5 MeV \\
  $V_\text{o}(\ell=0)$ & 66.717 MeV & 64.328 MeV \\
  $V_\text{o}(\ell=1)$ & 44.546 MeV &63.974 MeV \\
  $V_\text{o}(\ell=2)$ & 42.466 MeV &60.669 MeV \\
\end{tabular}
\end{ruledtabular}
\end{table}

The Furutani-Horiuchi-Tamagaki (FHT) finite-range two-body force is used as residual interaction~\cite{Furutani78,Furutani79} and possesses central, spin-orbit and tensor parts:
\begin{eqnarray}
V_{\rm C}(r) \!\!&=&\!\!  \sum_{n=1}^3  \, V_{\rm C}^n   \,    \left( W_{\rm C}^n + B_{\rm C}^n {\hat P}_{\sigma} - H_{\rm C}^n  {\hat P}_{\tau} - M_{\rm C}^n {\hat P}_{\sigma}{\hat P}_{\tau} \right)\: e^{-\beta_{\rm C}^n r^2}  \nonumber \\
V_{\rm LS}(r)  \!\!&=&\!\! \vec{L}\cdot \vec{S}\;\, \sum_{n=1}^2 \, V_{\rm LS}^n  \, \left( W_{\rm LS}^n - H_{\rm LS}^n  {\hat P}_{\tau}  \right) \, e^{-\beta_{\rm LS}^n r^2}   \nonumber \\
V_{\rm T}(r) \!\!&=&\!\!  S_{ij}\:\sum_{n=1}^3 \,  V_{\rm T}^n \, \left( W_{\rm T}^n  - H_{\rm T}^n  {\hat P}_{\tau} \right) \, r^2 e^{-\beta_{\rm T}^n r^2} ,
\label{eq.FHT123}
\end{eqnarray}
where $r\equiv r_{ij}$ is the  distance between the nucleons $i$ and $j$, $\vec{L}$ is the relative orbital angular momentum,  $\vec{S}=(\vec{\sigma}_i+\vec{\sigma}_j) / 2$ is the spin of the two particles,  $S_{ij}=3 (\vec{\sigma}_i \cdot \hat{r}) (\vec{\sigma}_j \cdot \hat{r})   - \vec{\sigma}_i \cdot \vec{\sigma}_j$ is the tensor operator, and ${\hat P}_{\sigma}$ and  ${\hat P}_{\tau}$ are spin and isospin exchange operators, respectively. $W$, $B$, $H$, and $M$ stand for the Wigner, Bartlett, Heisenberg, and Majorana parameters, respectively, which are dimensionless.
Each part of the interaction is the sum of two or three Gaussian radial form factors with different ranges: one for the  hard core, one to mimic the long range of the one-pion exchange potential, and one of intermediate range. The Coulomb interaction between valence protons is treated exactly (see Ref.~\cite{jaganathen_2017} for details).

Let us rewrite the interaction in terms of the spin-isospin projectors $\Pi_{ST}$ = $\Pi_{\textsf{t,t}}$, $\Pi_{\textsf{t,s}}$, $\Pi_{\textsf{s,t}}$, $\Pi_{\textsf{s,s}}$ \cite{jaganathen_2017}:
\begin{eqnarray}
V_{\rm C}(r) &=&  \nu^{\rm {C}}_{\textsf{t,t}} f^{\rm {C}}_{\textsf{t,t}}(r)  \Pi_{\textsf{t,t}} + \nu^{\rm {C}}_{\textsf{t,s}} f^{\rm {C}}_{\textsf{t,s}}(r) \Pi_{\textsf{t,s}} \nonumber  \\
&+& \nu^{\rm {C}}_{\textsf{s,s}} f^{\rm {C}}_{\textsf{s,s}}(r) \Pi_{\textsf{s,s}} + \nu^{\rm {C}}_{\textsf{s,t}} f^{\rm {C}}_{\textsf{s,t}}(r) \Pi_{\textsf{s,t}} \ ,   \nonumber  \\
V_{\rm LS}(r) &=&  \nu^{\rm {LS}}_{\textsf{t,t}} (\vec{L}\cdot \vec{S}) f^{\rm {LS}}_{\textsf{t,t}}(r) \Pi_{\textsf{t,t}} \ ,  \nonumber  \\
V_{\rm T}(r) &=& S_{ij} \left[\nu^{\rm {T}}_{\textsf{t,t}}f^{\rm {T}}_{\textsf{t,t}}(r) \Pi_{\textsf{t,t}} + \nu^{\rm {T}}_{\textsf{t,s}} f^{\rm {T}}_{\textsf{t,s}}(r) \Pi_{\textsf{t,s}}\right] \ ,
\label{eq.inter123}
\end{eqnarray}
where "$\textsf{s}$" and "$\textsf{t}$" stand for singlet ($S=0$ or $T=0$) and triplet ($S=1$ or $T=1$), respectively, and where coupling strengths
$\nu^{\rm {C}}_{\textsf{t,t}}$, $\nu^{\rm {C}}_{\textsf{t,s}}$, $\nu^{\rm {C}}_{\textsf{s,s}}$, $\nu^{\rm {C}}_{\textsf{s,t}}$, $\nu^{\rm {LS}}_{\textsf{t,t}}$, $\nu^{\rm {T}}_{\textsf{t,t}}$, and $\nu^{\rm T}_{\textsf{t,s}}$  are provided in spin-isospin channels.
The functions $f^{\rm {C}}(r)$, $f^{\rm {\rm {LS}}}(r)$ and $f^{\rm T}(r)$ in Eqs.~\ref{eq.inter123} are linear combinations of the original radial form-factors
appearing in Eqs.~\ref{eq.FHT123}. They are normalized to the first parameter $V_{\rm C}^1$, $V_{\rm LS}^1$ and $V_{\rm T}^1$, for central, spin-orbit, and tensor terms, respectively. More precisely, the constants $V_{\rm C}^{1,2,3}$, $V_{\rm LS}^{1,2}$ and $V_{\rm T}^{1,2}$ of Eqs.~\ref{eq.FHT123} are divided by $V_{\rm C}^1$, $V_{\rm LS}^1$ and $V_{\rm T}^1$, respectively, and multiplied afterwards by the associated coupling strength introduced in Eqs.~\ref{eq.inter123} for fixed values of $\textsf{s}$ and $\textsf{t}$. The remaining interaction parameters, that is the ranges of Gaussian form factors, their relative strength coupling constants, and the Wigner, Majorana, Bartlett, and Heisenberg parameters, bear their original values \cite{jaganathen_2017}. The fitted parameters of this interaction are adjusted to reproduce the energies of the low-lying states and proton separation energies of $^8$B and $^9$C (see Table~\ref{para2}).
\begin{table}[ht!]
\caption{Parameters of the FHT interaction in GSM and GSM-CC calculations. The superscripts C, LS, and T denote central, spin-orbit, and tensor, respectively. The indices "$\textsf{s}$" and "$\textsf{t}$" stand for singlet and triplet, respectively.
\label{para2}}
\begin{ruledtabular}
\begin{tabular}{cc}
  Parameter & Value \\
    \hline
  $\nu^{\rm {C}}_{\textsf{t,t}}$  & -1.669 MeV \\
  $\nu^{\rm {C}}_{\textsf{t,s}}$  & -5.204 MeV \\
  $\nu^{\rm {C}}_{\textsf{s,s}}$  & 1.233 MeV \\
  $\nu^{\rm {C}}_{\textsf{s,t}}$  & -4.019 MeV \\
  $\nu^{\rm {LS}}_{\textsf{t,t}}$  & -1427.795 MeV \\
  $\nu^{\rm {LS}}_{\textsf{s,t}}$  & 0 MeV \\
  $\nu^{\rm {T}}_{\textsf{t,t}}$  & 53.442 MeV fm$^{-2}$ \\
  $\nu^{\rm {T}}_{\textsf{t,s}}$  & -22.284 MeV fm$^{-2}$ \\
\end{tabular}
\end{ruledtabular}
\end{table}
They are optimized using the Gauss-Newton fitting procedure \cite{jaganathen_2017}.
Detailed discussions of the FHT interaction in the application of GSM can be found in Refs.~\cite{jaganathen_2017,Fossez15,dong17}.

Let us comment on the physical properties of the used FHT parameters of Table~\ref{para2}. By considering the singular values associated with the correlation matrix used in the Gauss-Newton fitting procedure, one can identify the well fitted and sloppy FHT parameters \cite{jaganathen_2017,GSMbook}. The singular values arise from the singular value decomposition of the correlation matrix
\cite{jaganathen_2017,GSMbook}. The singular values obtained in the fit of present FHT parameters are listed in Table~\ref{Table.SVD}.
\begin{table}[htb]
\begin{ruledtabular}
\caption{\label{Table.SVD} Singular values $s$ and their eigenvector components.  The largest components are depicted in boldface. See text for notations.}
\begin{tabular}{c|ccccccc}
$s$ & \phantom{$-$}$\nu^{\rm {C}}_{\textsf{t,t}}$ & \phantom{$-$}$\nu^{\rm {C}}_{\textsf{t,s}}$   & \phantom{$-$}$\nu^{\rm {C}}_{\textsf{s,s}}$ & \phantom{$-$}$\nu^{\rm {C}}_{\textsf{s,t}}$  & \phantom{$-$}$\nu^{\rm {LS}}_{\textsf{t,t}}$  & \phantom{$-$}$\nu^{\rm {T}}_{\textsf{t,t}}$ & \phantom{$-$}$\nu^{\rm {T}}_{\textsf{t,s}}$     \\ \hline\\[-7pt]
2.582               & \phantom{$-$}0.01  & \textbf{\phantom{$-$}0.83} & $-$0.03            & \textbf{\phantom{$-$}0.55}  & \phantom{$-$}0.00  & $-$0.02           & \phantom{$-$}0.02        \\
0.277               & \phantom{$-$}0.03  & \textbf{$-$0.55}            & $-$0.07            & \textbf{\phantom{$-$}0.83}  & \phantom{$-$}0.00  & $-$0.05           & $-$0.04                  \\
0.051               & $-$0.03            & $-$0.04            & \phantom{$-$}0.07  & \phantom{$-$}0.03  & \phantom{$-$}0.00  & \phantom{$-$}0.07 & \textbf{\phantom{$-$}0.99}        \\
0.028               & \textbf{\phantom{$-$}0.81}  & \phantom{$-$}0.00  & \textbf{\phantom{$-$}0.46}  & $-$0.01            & \phantom{$-$}0.00  & \textbf{$-$0.37}           & \phantom{$-$}0.02        \\
0.017               & \textbf{$-$0.45}            & $-$0.01            & \textbf{\phantom{$-$}0.88}  & \phantom{$-$}0.09  & \phantom{$-$}0.00  & \phantom{$-$}0.10 & $-$0.08                  \\
0.001 & \textbf{\phantom{$-$}0.38}  & \phantom{$-$}0.00  & \phantom{$-$}0.08  & \phantom{$-$}0.05  & \phantom{$-$}0.01  & \textbf{\phantom{$-$}0.92} & $-$0.06                  \\
$\sim 0$   & \phantom{$-$}0.00  & \phantom{$-$}0.00  & \phantom{$-$}0.00  & \phantom{$-$}0.00  & \textbf{$\sim$1.00}  & $-$0.01           & $-$0.00
 \end{tabular}
\end{ruledtabular}
\end{table}
One can see that the $\nu^{\rm {C}}_{\textsf{t,s}}$ and $\nu^{\rm {C}}_{\textsf{s,t}}$ parameters are well defined, as they bear the largest singular values. Conversely, $\nu^{\rm {LS}}_{\textsf{t,t}}$ and $\nu^{\rm {T}}_{\textsf{t,t}}$ are the most sloppy parameters, as their singular values are close to zero. Remaining parameters, i.e.~$\nu^{\rm {C}}_{\textsf{t,t}}$, $\nu^{\rm {C}}_{\textsf{s,s}}$ and $\nu^{\rm {T}}_{\textsf{t,s}}$, are mildly important as their associated singular values lie in between. This is consistent with the analysis done in Ref.\cite{jaganathen_2017}, where fitted FHT parameters  possess similar statistical properties. Consequently, even though the FHT interaction had to be refitted as the model spaces of Ref.\cite{jaganathen_2017} and of this paper are different, both interactions can be considered to be of the same quality from a physical point of view.

The considered GSM model space for protons consist of the $psd$ partial waves. The proton $0p_{3/2}$ partial wave is represented by the Berggren ensemble, with the resonant s.p.~state $0p_{3/2}$ and 30 s.p.~states in the non-resonant continuum along the contour $\mathcal{L}^+_{p_{3/2}}$. This contour consists of three segments connecting the points: $k_{\text{min}}$=0.0, $k_{\text{peak}}=0.15-i0.10$ fm$^{-1}$, $k_{\text{middle}}$=0.3 fm$^{-1}$ and $k_{\text{max}}$=2.0 fm$^{-1}$, and each segment is discretized with 10 points. The $p_{1/2}$,  $d_{3/2}$ and $d_{5/2}$ partial waves are represented with six scattering-like harmonic oscillator (HO) states, while for the $s_{1/2}$ partial wave, five scattering-like HO states are considered. A scattering-like (resonant-like) HO state is treated as a scattering (resonant) state for model space truncation, that is when particle-hole excitations to the non-resonant continuum are imposed. Each s.p.~state of the Berggren ensemble indeed becomes a valence shell in the many-body GSM and GSM-CC calculation, so that model spaces must be truncated in order for Hamiltonian matrix dimensions to be tractable.
For the neutron space, only $0p_{3/2}$ and $0p_{1/2}$ resonant-like HO s.p.~states, as well as three scattering-like HO states in the $1s_{1/2}$, $0d_{5/2}$ and $0d_{3/2}$ partial waves, are included.

We recall that the $0s_{1/2}$ shells do not belong to the model space as they are fully occupied and form the $^4$He core. Thus, the GSM and GSM-CC calculations are performed in 54 shells for protons: 31 $p_{3/2}$ shell, 23 $0p_{1/2}$, $s_{1/2}$, $d_{3/2}$, and $d_{5/2}$ scattering-like shells, and 5 shells for neutrons: 2 resonant-like shells $p_{3/2}$ and $p_{1/2}$, 3 scattering-like shells $s_{1/2}$, $d_{3/2}$, and $d_{5/2}$.

The two-body GSM-CC Hamiltonian is usually re-scaled by multiplicative corrective factors $c(J^{\pi}) \sim 1$ for $J^{\pi}=3/2^-_1, 1/2^-_1, 5/2^-_1$ states. They are used to compensate for missing channels in the GSM-CC wave function, such as the non-resonant channels induced by the many-body continuum states of $^8$B, channels involving many-body projectiles, etc. The corrective factors used in this work are: $c(3/2^-)=1.0155$,  $c(1/2^-)=1.0958$, $c(5/2^-)=1.035$.

\subsection{Energy spectrum of $^8$B and $^9$C}
\label{sec3-a}
In the GSM-CC calculation of the $^8$B$(p,\gamma)$$^9$C reaction observables, we firstly calculate the low-lying states of the target nucleus $^8$B with GSM. The channel states in GSM-CC are then built by coupling three lowest states of the spectrum of $^8$B, i.e., the ground state $J^\pi=2^+_1$, the first and the second excited states, $J^\pi=1^+_1$ and $J^\pi=3^+_1$ (which are both low-lying resonances) with the proton in partial waves: $s_{1/2}$, $p_{1/2}$, $p_{3/2}$, $d_{3/2}$ and $d_{5/2}$. While no truncation had to be imposed in the GSM model space of $^8$B eigenstates, we had to truncate the GSM-CC model space of $^9$C for matrix dimensions to be numerically tractable therein: at most three nucleons are allowed to occupy scattering and scattering-like shells in the basis of Slater determinants.
\begin{figure}[htb]
\includegraphics[width=0.9\linewidth]{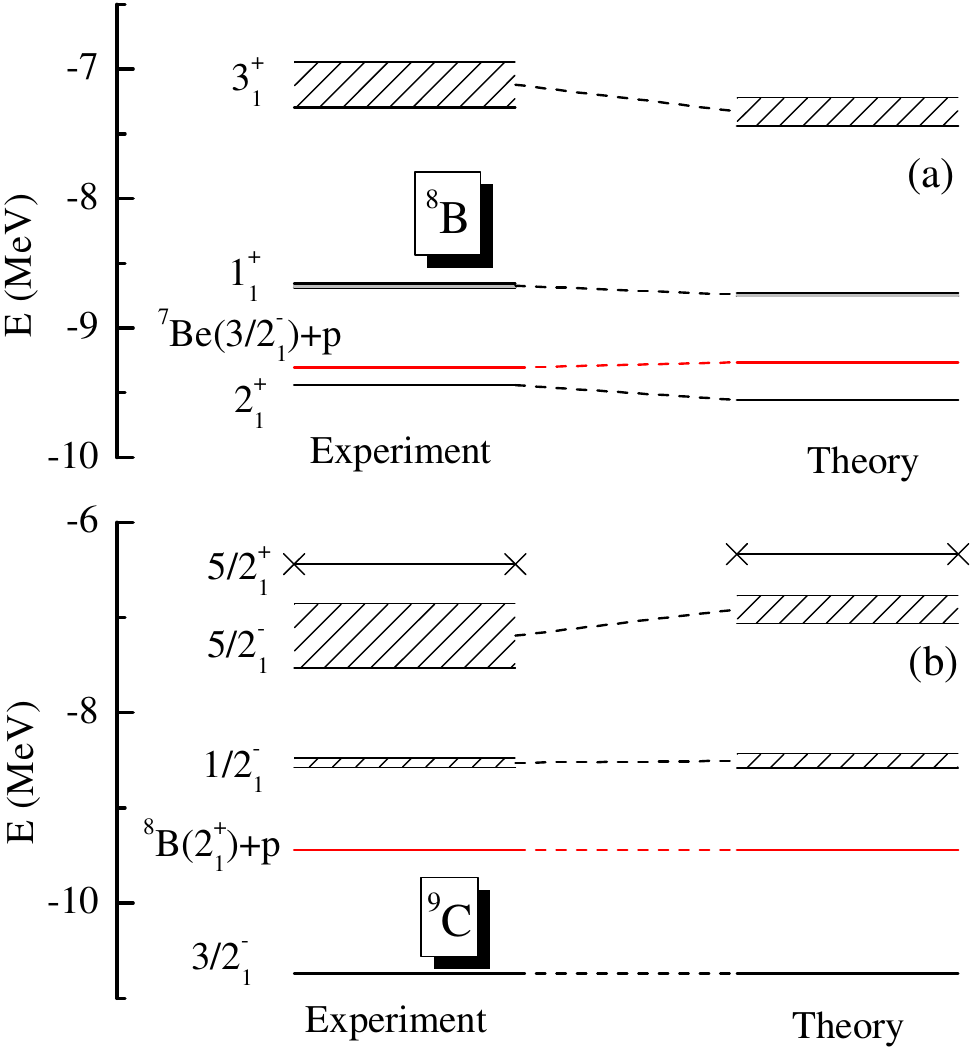}
\caption{The GSM energy levels of $^8$B and GSM-CC energy levels of $^9$C are compared with the experimental data~\cite{nndc,Brown17} in panels (a) and (b), respectively. Energies are given relatively to that of the $^4$He core. The measured ${5/2}^+_1$ state has a very broad width, which is not shown in the figure. For more details see Table \ref{spectra} and discussion in the text.}
\label{fig-1}
\end{figure}

\begin{table}[ht!]
\caption{The calculated excitation energies of low energy bound and resonance states in $^{8}$B and $^{9}$C are compared with the experimental data. All energies are given relatively to their respective ground states. Experimental data are taken from Refs.~\cite{nndc,Brown17}.
\label{spectra}}
\begin{ruledtabular}
\begin{tabular}{c|c|c|c|c|c|c}
 &\multicolumn{3}{c|}{Theory}& \multicolumn{3}{c}{Experiment} \\
   \hline
  & J$^{\pi}$ & E(MeV) & $\Gamma$(keV) &  J$^{\pi}$ &   E(MeV) & $\Gamma$(keV)  \\
    \hline

  $^{8}$B & ${2}^+_1$  &0.000 & - &  ${2}^+_1$& 0.000& -\\
  &  ${1}^+_1$  & 0.891 & 27.90 &${1}^+_1$ &0.7695(25) & 35.6 (6) \\
 &   ${3}^+_1$  & 2.149 &280.51& ${3}^+_1$ & 2.32(20)& 350 (30)\\
    \hline
$^{9}$C & ${3/2}^-_1$  &0.000 & - &  ${3/2}^-_1$& 0.000& -\\
  &      ${1/2}^-_1$  &2.231 & 143.62 &  ${1/2}^-_1$& 2.218(11)& 52(11)\\
   &     ${5/2}^-_1$  &3.822 & 297.91 &  ${5/2}^-_1$& 3.549(20)& 673(50)\\
   &    ${3/2}^+_1$  &4.361 & 323.398 &  ${5/2}^+_1$\footnote{The latest measured energy of ${5/2}^+_1$ is 4.3 (3) MeV, and width is $4.0^{+2.0}_{-1.4}$ MeV ~\cite{Hooker19}. }& 4.40(4)& 2750(110)\\
   &    ${5/2}^+_1$  &4.395 & 327.96 &  & & \\
   &     ${1/2}^+_1$  &5.067 & 359.231 &   ? & 5.75(4) & 601(50)\\
  \end{tabular}
\end{ruledtabular}
\end{table}

The GSM energies and widths of the low-lying states of $^8$B are given in Fig.~\ref{fig-1}(a), together with experimental data. The calculated ground-state energy of $^8$B with respect to $^4$He is -9.609 MeV, which is close to the experimental value -9.441 MeV~\cite{nndc}. The calculated energies and widths of the first and second excited states in $^8$B are also very satisfactory. The GSM-CC energies and widths of $^9$C states as well as proton separation energy are depicted in Fig.~\ref{fig-1}(b), where one can see that experimental data are also very well reproduced. Energy levels and proton-emission widths are also listed in Table.~\ref{spectra} for $^8$B and $^9$C along with experimental data.

As seen in the figure, the first excited state of $^9$C is a resonance. The experimental positive parity state $5/2^+_1$ is a very broad resonance, with the reported width $\Gamma=$2.750 (110) MeV in Ref. \cite{Brown17}, and $\Gamma=4.0^{+2.0}_{-1.4}$ MeV in the more recent measurement~\cite{Hooker19}. In GSM-CC calculation, the energy of $5/2^+_1$ state is close to the experimental value but its width is smaller than in experimental data. The energy of the positive parity state, $3/2^+_1$, is very close to that of $5/2^+_1$. The resonance with unknown spin and parity in experimental data is predicted to be the ${1/2}^+_1$ state. It is clear that the calculated energies of low-lying states in $^9$C are close to experimental data.

\subsection{The astrophysical factor for $^8$B$(p,\gamma)$$^9$C reaction}
\label{sec3-b}

To calculate the antisymmetrized matrix elements of the electromagnetic operators in GSM-CC, we use the method described in Ref.\cite{Fossez15}. For this, electromagnetic operators are separated in short-range and long-range parts \cite{Fossez15,dong17,GSMbook}. Short-range operators are handled with a HO expansion as they are important only in the nuclear zone. Long-range operators, which extend in the asymptotic region, are considered with complex rotation, so that the range of radial integrals therein is infinite \cite{Fossez15,dong17,GSMbook}. Consequently, the effect of the infinite-range of electromagnetic operators is fully taken into account in the radial integrals of GSM-CC matrix elements.

In order for radiative capture reaction cross sections to be very precisely computed in GSM-CC, we introduce the experimental ground-state energy of $^8$B in the coupled-channel equations instead of its value provided by GSM. This ensures that the experimental proton separation energy in $^9$C is exactly taken into account in the cross section calculation. This is necessary due to the very small proton separation energy of $^8$B, of the order of 100 keV.
\begin{figure}[htb]
\includegraphics[width=0.9\linewidth]{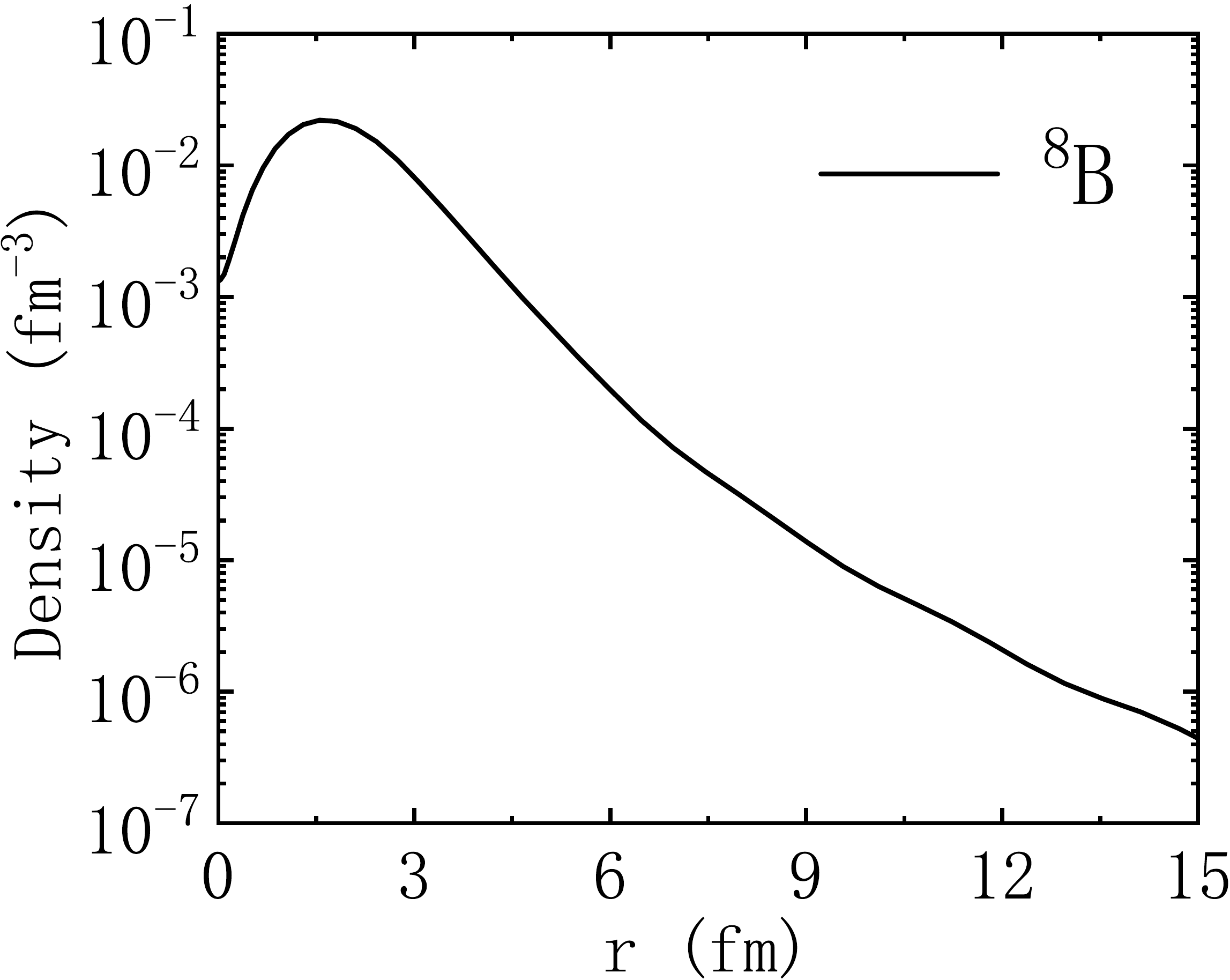}
\caption{The calculated GSM one-body density of valence protons in the ground state ${2}^+_1$ of $^8$B.}
\label{fig:density}
\end{figure}
Added to that, the halo structure of the $^8$B ground state~\cite{B8halo1,B8halo2} must be well reproduced in GSM for GSM-CC cross sections to be reliable. In order to assess the asymptotic behavior of the ground state of $^8$B, its proton one-body density has been calculated in GSM (see Fig.~\ref{fig:density}). In our GSM calculation, a very extended density distribution develops, which is consistent with the proton halo seen in experimental data. Indeed, one can see that the one-proton density of $^8$B slowly decreases and is typical of a weakly bound many-body state. The calculated root mean square (RMS) proton radius in GSM is 2.63 fm, which is close to the experimental value of 2.76(9) fm~\cite{B8halo2}. This proves that the nuclear asymptotes of $^8$B target states are well reproduced and hence that the proton capture cross section of $^8$B in the low center of mass (c.m.) energy region can be precisely calculated. Note that the maximal radius of 15 fm is arbitrary and is proper to Fig.~\ref{fig:density}. The radial matrix elements entering cross sections in GSM-CC always extend to infinity (see discussion above).

\begin{table}[ht!]
\caption{{The electric quadrupole and magnetic moments of the ground state of $^{8}$B and $^{9}$C obtained with GSM and compared to standard shell model (SM) and experimental data. The experimental electric quadrupole moment of $^{8}$B is taken from Ref.~\cite{Stone16} and the experimental magnetic moments of $^{8}$B and $^{9}$C from Ref.~\cite{Tilley04} and Ref.~\cite{Huhta98}, respectively. The GSM calculations of electric quadrupole moment are performed with different effective charges. For SM results, we show results of Huhta et al.~\cite{Huhta98}. Other SM calculations are done using OXBASH code~\cite{oxbash} and the same $p$-shell interaction~\cite{ptbme} as in Ref.~\cite{Huhta98}. For more details, see description in the text.} }
\label{q-moment}
\begin{ruledtabular}
\begin{tabular}{lcc}
  & $^{8}$B,~${2}^+_1$ & $^{9}$C,~${3/2}^-_1$\\
    \hline
  $Q\left(e^2 \textrm{fm}^2 \right)$  &  &  \\
  Experiment  & +6.43(14) &  \\
  GSM~(1.00$e$,~0.00$e$)  & +1.622  &  -0.748 \\
  GSM~(1.26$e$,~0.47$e$)~\cite{Hees1988}   & +3.083 & -2.424  \\
  GSM~(recoil correction) & +1.516 &-0.871 \\
  SM(~(1.00$e$,~0.00$e$)  & +1.997  &-1.401   \\
  SM~(1.26$e$,~0.47$e$)~\cite{Hees1988}   & +2.970  &-2.593   \\
  SM~(recoil correction) & +1.729 &-1.324 \\
\hline
 $\mu \left(\mu_\textsc{N} \right)$  &  &  \\
  Experiment  & 1.0355(3) & (-)1.396(3)\\
  GSM  & 0.943  & -1.173  \\
  SM  & 1.132  & -1.438 \cite{Huhta98} \\
\end{tabular}
\end{ruledtabular}
\end{table}

We will then focus on the study of proton capture cross section reaction $^8$B(p,$\gamma$)$^9$C with GSM-CC. For this, the proton capture cross section to the final state of $^9$C with total angular momentum ${ {J}_{f} }$ is obtained from the angular integration of differential cross-sections. The matrix elements of the electromagnetic operators between the antisymmetrized initial and final states of $^8$B and $^9$C are used for that matter. Electromagnetic transitions connect the scattering states of $^9$C with the capturing state in $^9$C, which can be either the bound state $J_f = {3/2}_1^-$ or the resonance $J_f = {1/2}_1^-$. We consider scattering composite entrance states p + $^8$B with $J_i = {1/2}^+,~{3/2}^+,~{5/2}^+$ for the E1 transition, $J_i = {1/2}^-,~{3/2}^-,~{5/2}^-$ for the M1 transition and $J_i = {1/2}^-,~{3/2}^-, ~ {5/2}^-,~{7/2}^-$ for the E2 transition. The total cross section is them the sum of all the possible cross section terms function of $J_i$ and $J_f$ incoming and capturing states \cite{Fossez15,dong17,GSMbook}.

To remove the exponential dependence on energy of the cross section arising from the Coulomb barrier, it is convenient to consider the astrophysical $S$ factor instead of the radiative capture cross section for charged particles:
\begin{equation}
S(E_{\rm c.m.})={ \sigma }(E_{\rm c.m.})E_{\rm c.m.}e^{2\pi\eta}\; .
	\label{eq_S}
\end{equation}
$\eta$ in this expression is the Sommerfeld parameter $\eta=e^2 Z_1 Z_2 \mu_{p}/ \hbar^2 k$, with $\mu_{p}$ the proton effective mass (see Eq.~\ref{GSM_Hamiltonian}).

For the calculation of E1 and E2 transitions, effective charges are used. For E1 transitions, the empirical effective charges~\cite{Hornyak75,YKHo88} are:
\begin{equation}
	{ {e}_{ \text{eff} }^{p}({\rm E1}) = e \left( 1 - \frac{Z}{A} \right) \ ; }\qquad
    {e}_{ \text{eff} }^{n}({\rm E1}) = - e f_{\rm E1} \frac{Z}{A} \; ,
	\label{eq1}
\end{equation}
where ${ Z }$ and ${ A }$ are the proton number and total particle number in the combined system of a target nucleus and a projectile. The effective charges originate from the recoil corrections of the center of mass only~\cite{Hornyak75,YKHo88}.
For E2 transitions, the commonly used values of the proton and the neutron effective charges are in the range of $1.1 e < {e}_{ \text{eff} }^{p}({\rm E2}) < 1.5 e$ and $0.5 e < {e}_{ \text{eff} }^{n}({\rm E2}) < 0.6 e$, respectively~\cite{Prestwich84,Castel86,Hees1988,Rydt2009}. No effective charge enters M1 transitions.

In Table~\ref{q-moment}, we show the electric quadrupole and magnetic moments of $^{8}$B and $^{9}$C ground states obtained with GSM. Electric quadrupole moments are calculated with different values of effective charges. $(e_p, e_n) = (1.26$e$,~0.47$e$)$ were adopted in Ref.~\cite{Hees1988} when calculating the energy levels and static moments of 0$p$-shell nuclei in standard shell model (SM). The effective charges arising from the recoil of the center of mass read~\cite{Hornyak75,YKHo88}:
\begin{equation}
{ {e}_{ \text{eff} }^{p}({\rm E2}) = e \left( 1 - \frac{2}{A} + \frac{Z}{ {A}^{2} } \right) \ ; }\qquad
{e}_{ \text{eff} }^{n}({\rm E2}) = Z/A^2 e  \; ,
\label{e2effcharge}
\end{equation}
 and are labeled ``recoil correction''. One may notice that a large effective charge would be needed to reproduce the experimental quadrupole moment of $^8$B.

\begin{figure}[htbp]
	\includegraphics[width=0.9\linewidth]{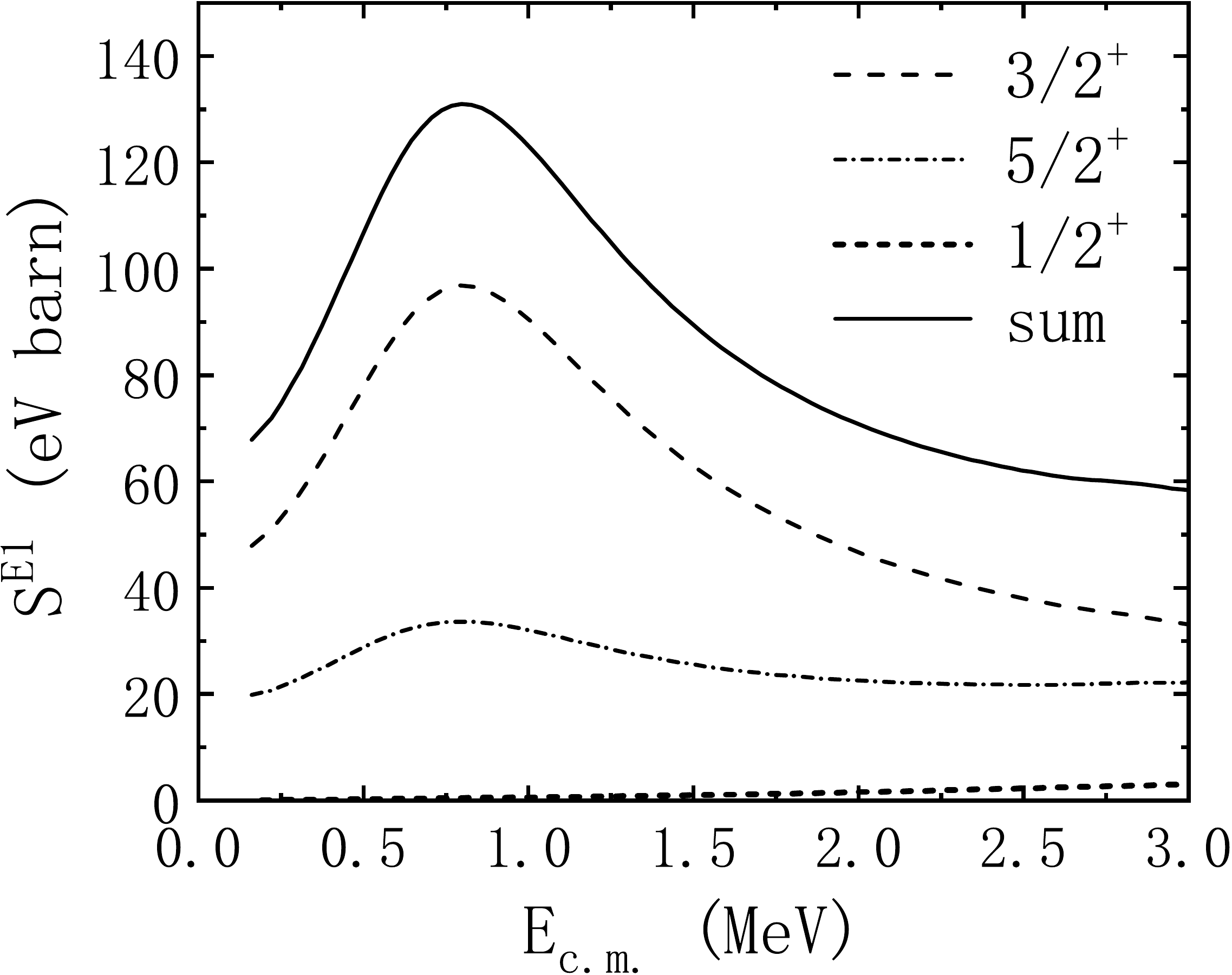}
	\caption{The E1 astrophysical factor of the $^8$B$(p,\gamma)$$^9$C reaction is plotted as a function of proton projectile energy in the p + $^8$B center of mass frame. The solid line represents the fully antisymmetrized GSM-CC calculation for the radiative proton capture to the ground state of $^9$C. The dashed, dashed-dotted, and dotted lines show the contributions from the radiative proton capture to the $^9$C ground state from the initial p + $^8$B composite scattering states with $J_i^{\pi}=$ 3/2$^+$, 5/2$^+$, and 1/2$^+$, respectively. }
	\label{fig-2}
\end{figure}
\begin{figure}[htbp]
	\includegraphics[width=0.9\linewidth]{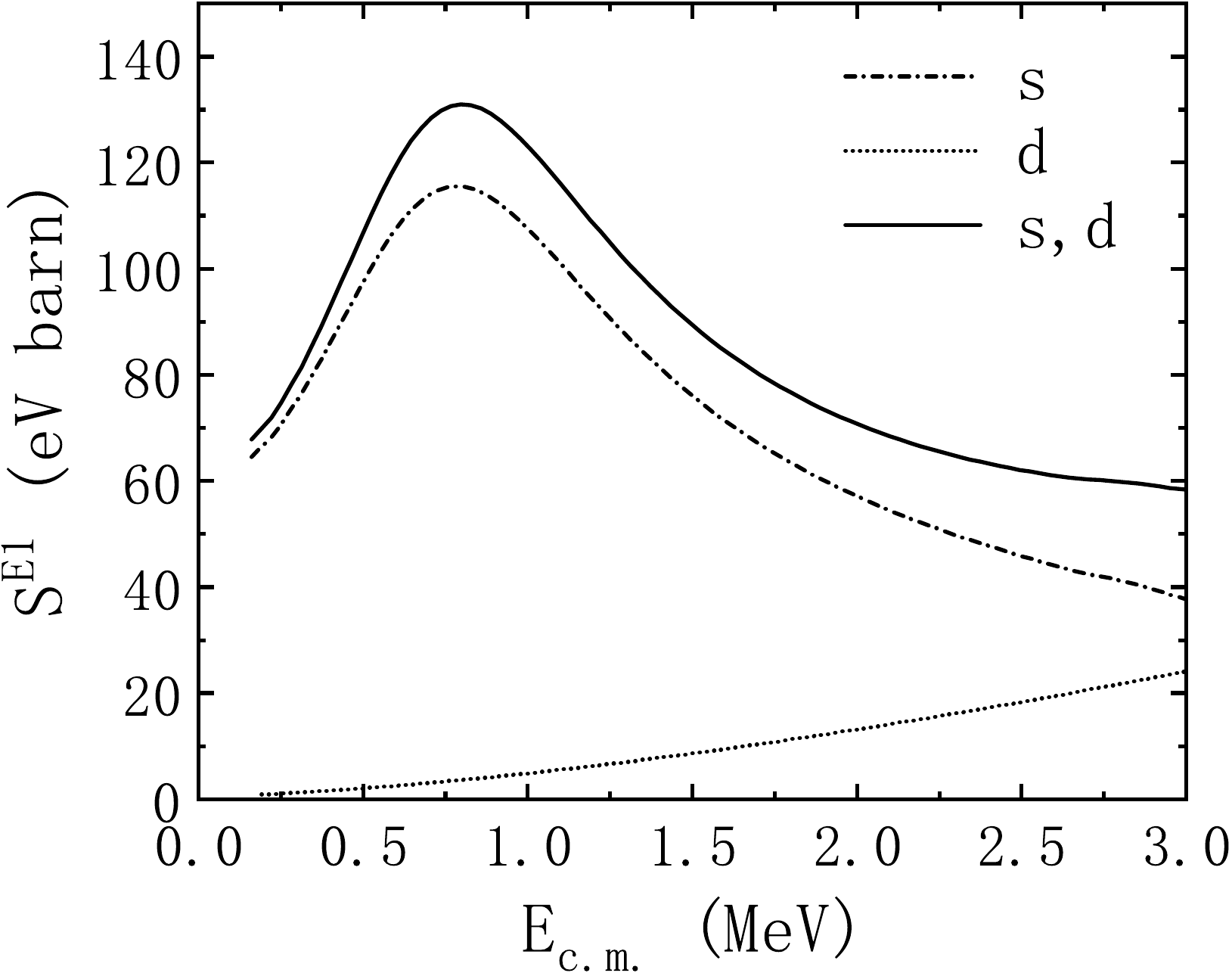}
	\caption{The solid line shows the E1 astrophysical factor $S^{\rm E1}$ as a function of proton projectile energy in the p + $^8$B center of mass frame for the proton radiative capture to the ground state of $^9$C. The E1 radiative capture of $s$- and $d$-wave protons is shown as the dashed-dotted and dotted lines, respectively. }
	\label{fig-3}
\end{figure}
\begin{figure}[htbp]	
	\includegraphics[width=0.9\linewidth]{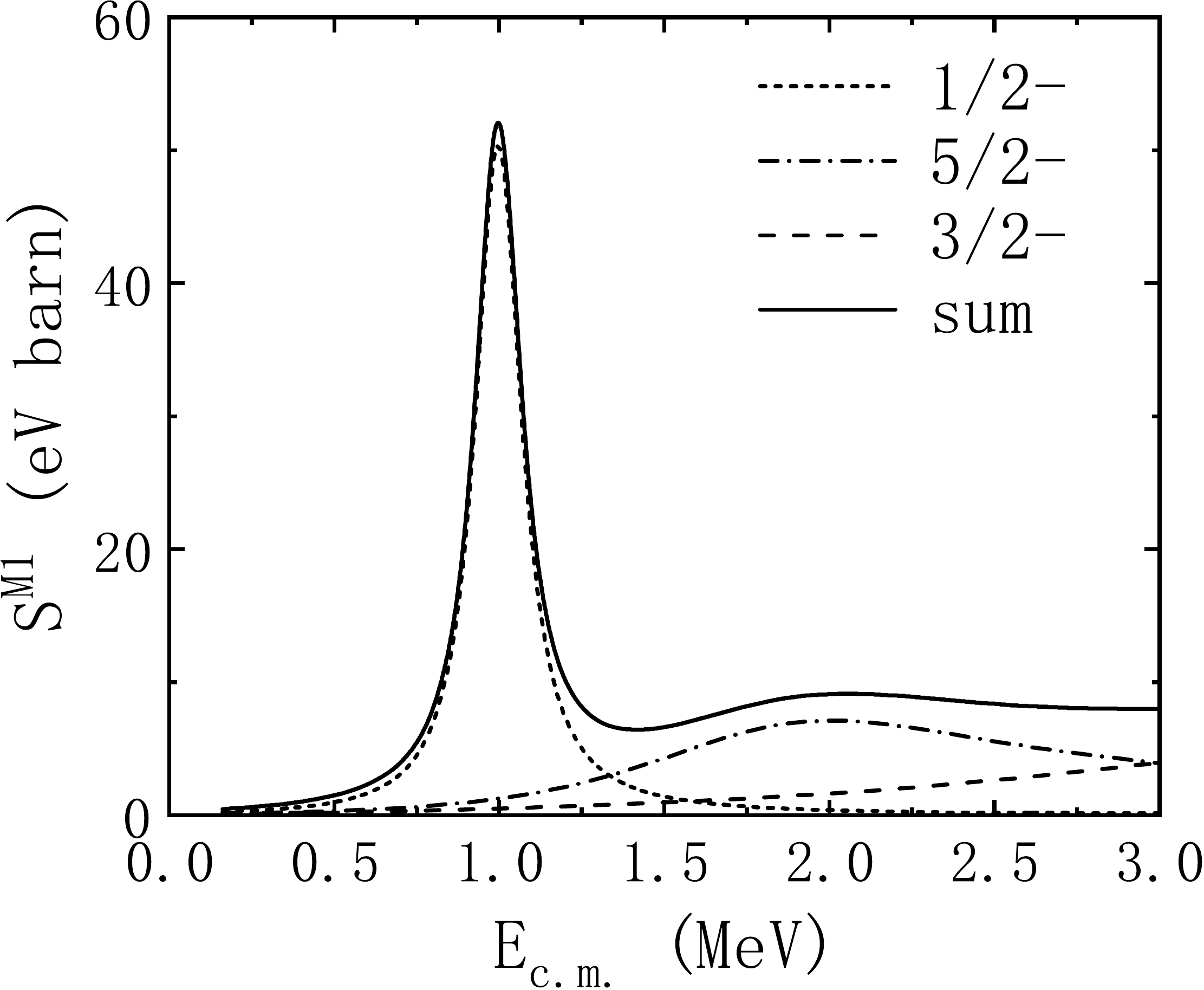}
	\caption{The M1 astrophysical factor for the radiative proton capture to the ground state of $^9$C as a function of proton projectile energy is plotted (solid line). The peak corresponds to the ${1/2}_{1}^-$ resonance of $^9\text{C}$. The dashed, dashed-dotted, and dotted lines show the contributions from the radiative proton capture to the $^9$C ground state from the initial p + $^8$B composite scattering states with $J_i^{\pi}$=1/2$^-$, 3/2$^-$, and 5/2$^-$,  respectively.
}
	\label{fig-4}
\end{figure}
\begin{figure}[htbp]
	\includegraphics[width=0.90\linewidth]{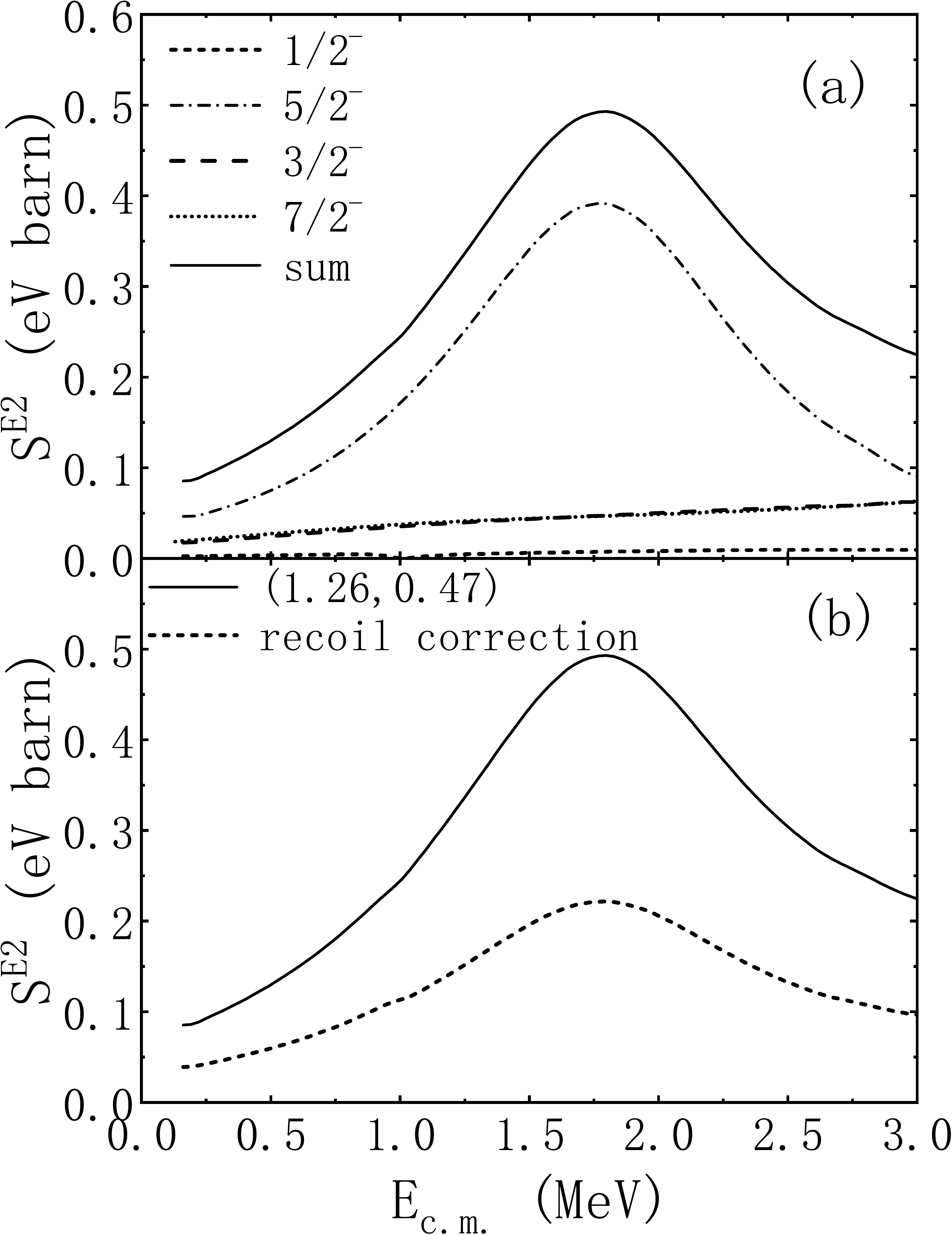}
	\caption{The E2 astrophysical factor is depicted as a function of proton projectile energy. In the upper panel (a), the solid line shows the sum of individual contributions of different initial p + $^8$B composite scattering states  $J_i^{\pi}=3/2^-_1,~1/2^-_1,~5/2^-_1$ and $7/2^-_1$ to the astrophysical $S^{\rm E2}$ factor of the radiative proton capture to the ground state of $^9$C. The E2 effective charges $(e_{ \text{eff} }^p, e_{ \text{eff} }^n) = (1.26$e$,~0.47$e$)$~\cite{Hees1988} are used. The lower panel (b) compares two different sets of E2 effective charges. The solid line is the same as in the panel (a). The dotted line is calculated for the E2 effective charges issued from the recoil of the center of mass (Eq. \ref{e2effcharge}).}
		\label{fig-5}
\end{figure}

\begin{figure}[htbp]
	\includegraphics[width=0.9\linewidth]{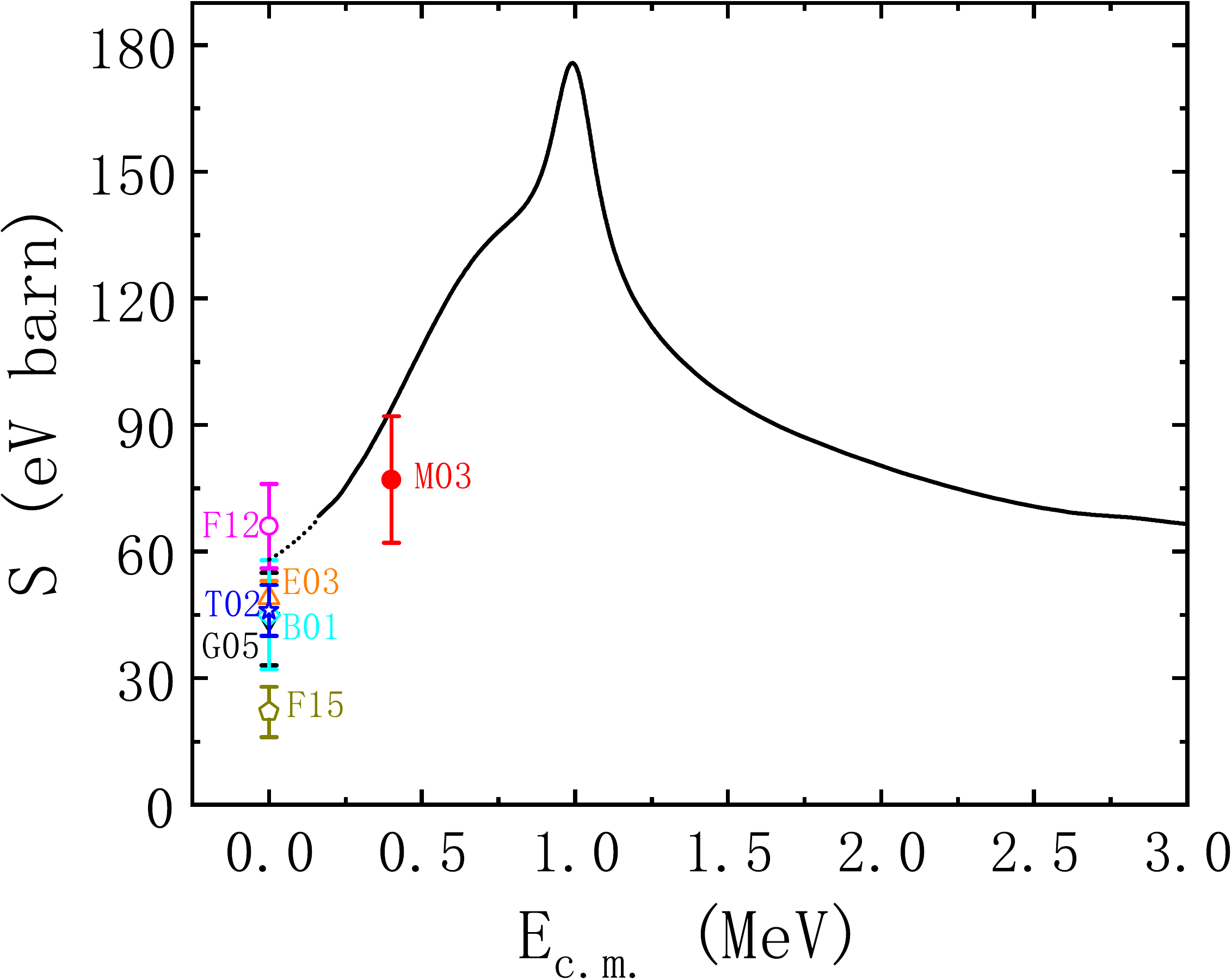}
	\caption{
The total astrophysical $S$ factor for the radiative proton capture to the ground state of $^9$C is plotted as a function of proton projectile energy. The astrophysical factor of the proton radiative capture reaction in the limit $E_{\rm c.m.}\rightarrow 0$ has been extracted by using an expansion from a quadratic polynomial (dashed line), whose parameters are obtained by fitting the calculated function $S(E_{\rm c.m.})$ in the energy range 0.1$\leq E_{\rm c.m.}\leq $0.3 MeV. The experimental data for the astrophysical factor in the limit $E_{\rm c.m.}\rightarrow 0$, taken from Refs.~\cite{Enders03,Trache02,Beaumel01,BGuo05,Fukui12,Fukui15}, are labeled as ``E03'', ``T02'', ``B01'', ``G05'', ``F12'', and ``F15'', respectively. The astrophysical factor extracted from the Coulomb dissociation experiment~\cite{Moto03} in the energy range 0.2 MeV $\leq E_{\rm c.m.} \leq$ 0.6 MeV is labeled as ``M03''.}
	\label{fig-6}
\end{figure}
\begin{figure}[htb]
\includegraphics[width=0.9\linewidth]{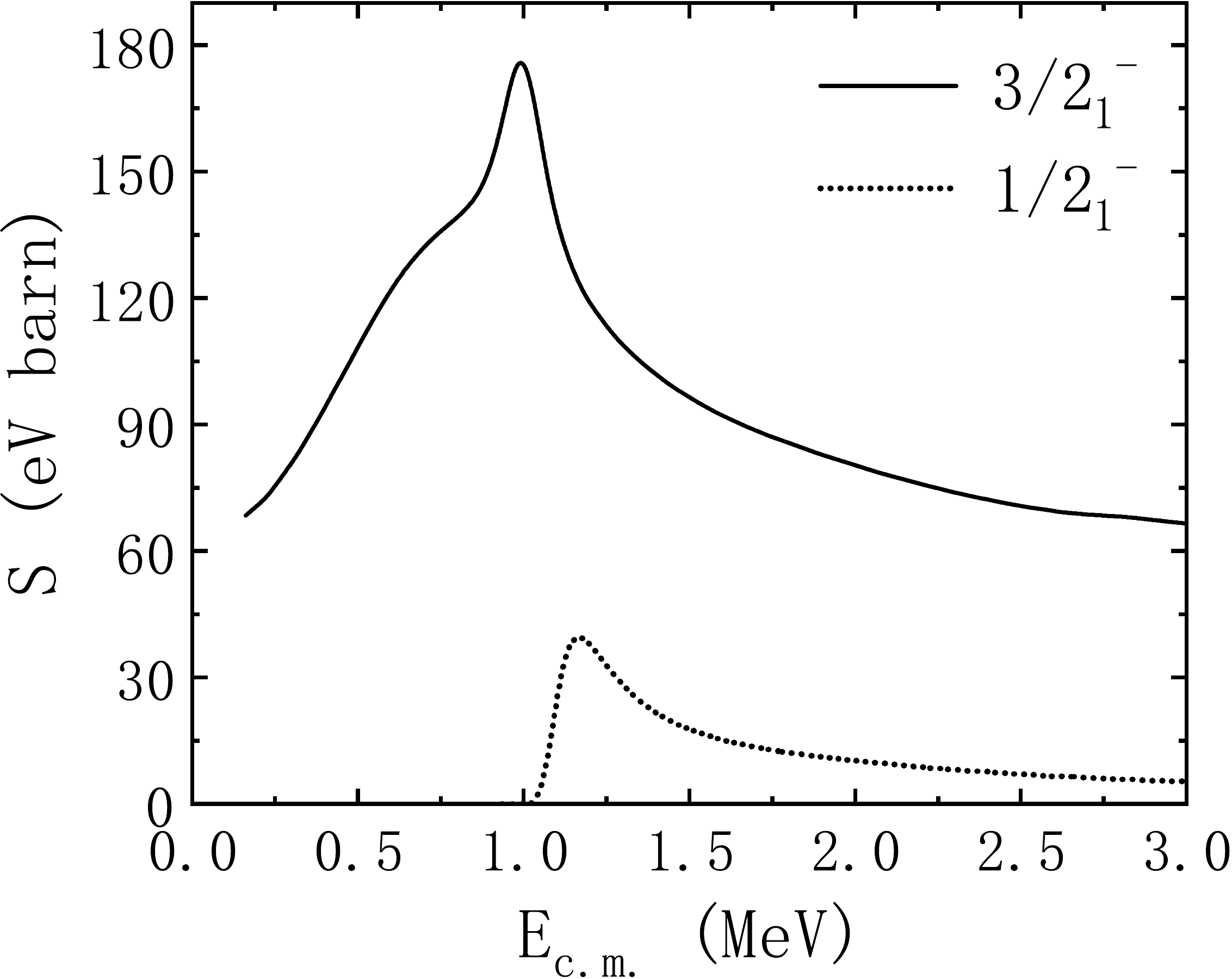}
\caption{Calculation of the total astrophysical factor of the $^8$B$(p,\gamma)$$^9$C reaction with GSM-CC. The capture reaction to the ground state of $^9$C is shown by the solid line, while the capture to the first excited resonance state, 1/2$^-_1$, is provided by the dotted line.  }
\label{fig-7}
\end{figure}
\begin{figure}[htbp]
	\includegraphics[width=0.9\linewidth]{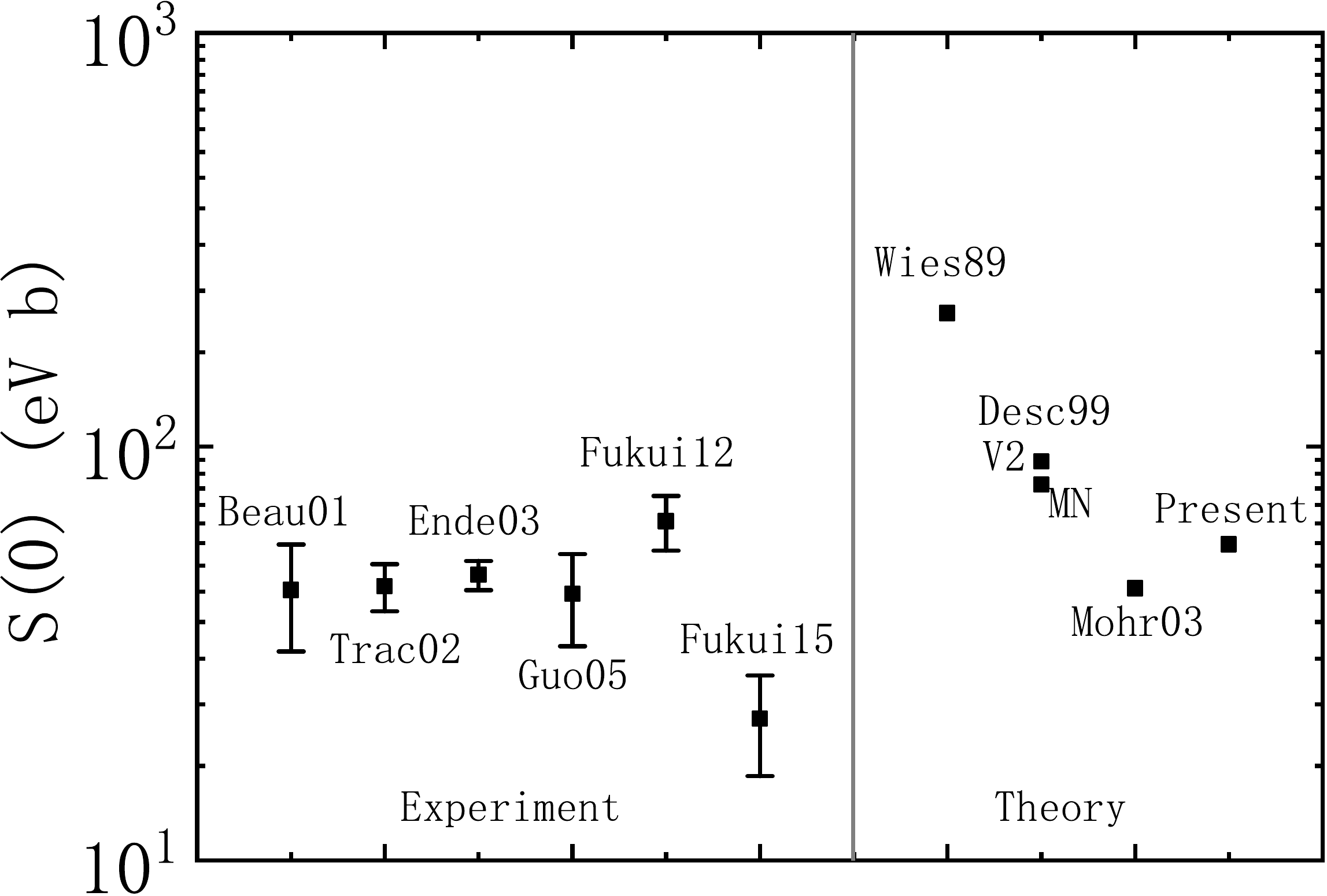}
	\caption{Comparison between different experimental analyses~\cite{Enders03,Trache02,Beaumel01,BGuo05,Fukui12,Fukui15} and various theoretical calculations~\cite{Wiescher89,Descouvemont99,Mohr03} of the total astrophysical $S$ factor at $E_{\rm c.m.}=0$. The GSM-CC result is labeled as ``Present''. }
	\label{fig-6b}
\end{figure}

Figs.~\ref{fig-2}\--\ref{fig-5} show separate contributions to the total astrophysical $S$ factor for the  $^8$B$(p,\gamma)$$^9$C reaction to the ground state of $^9$C: $S^{\rm E1}$ for E1 transitions (Figs.~\ref{fig-2} and ~\ref{fig-3}); $S^{\rm M1}$ for M1 transitions (Fig.~\ref{fig-4}); $S^{\rm E2}$ for E2 transitions (Fig.~\ref{fig-5}). The major contribution to the $S$ factor comes from the E1 transitions. Especially, the capture of s-wave proton to the ground state of $^9$C plays a crucial role, as seen in Fig.~\ref{fig-3}. This is also consistent with results from previous studies~\cite{Trache02,Mohr03,BGuo05}.
One can see in Fig.~\ref{fig-2} that the values of $S^{E1}$ first increase with ${E}_{ \text{c.m.} }$ and then decrease, forming a broad bump at
${E}_{ \text{c.m.} }\sim 0.6$ MeV.
The largest contribution to $S^{E1}$ comes from the initial continuum states $J_i^{\pi}=$ 3/2$^+$, 5/2$^+$ which both exhibit a broad peak at around 0.6 MeV. The contribution of the $J_i^{\pi}=$ 3/2$^+$ continuum is larger than that from the $J_i^{\pi}=$5/2$^+$.  A similar conclusion was drawn in the studies performed with the microscopic cluster model~\cite{Descouvemont93,Descouvemont99}.

In Fig.~\ref{fig-4} one can see a sharp peak for the M1 transition, which is caused by the ${ {1/2}_{1}^{-} }$ resonance. The ${ {1/2}_{1}^{-} }$ resonance indeed lies above the one-proton decay threshold. The center-of-mass energy of this peak is: ${E}_{ \text{c.m.} } = {E}_{i}^{ ( A ) } [ \text{GSM-CC} ] - {E}_{0}^{ ( A - 1 ) } [ \text{GSM} ]$, where ${ {E}_{i}^{ ( A ) } [ \text{GSM-CC} ] }$ is the GSM-CC energy of resonance ${ i }$ in $^9$C, and ${ {E}_{0}^{ (A-1) } [ \text{GSM} ] }$ denotes the GSM ground-state energy of $^8$B. $S^{M1}$ is fairly small at low energies, but its contribution becomes comparable to $S^{E1}$ in the energy region of the ${1/2}_1^-$ resonance.

$S^{\rm E2}$ is significantly smaller than both $S^{\rm E1}$ and $S^{\rm M1}$, as seen in Fig.~\ref{fig-5}. The ${ {5/2}_{1}^{-} }$ resonance plays an important role in $S^{\rm E2}$, while the contribution of the ${ {1/2}_{1}^{-} }$ resonance is small. A similar conclusion was made in the cluster model~\cite{Descouvemont99}.

The total astrophysical factor $S = S^{E1} + S^{M1} +S^{E2}$ is shown in Fig.~\ref{fig-6}. For a standard value of the effective charge for $E1$ and $E2$, the astrophysical factor calculated in the GSM-CC is close to most of existing data. Experimental values for the astrophysical $S$ factor in the limit $E_{\rm c.m.}\rightarrow$ 0~\cite{Enders03,Trache02,Beaumel01,BGuo05} seem to be smaller than those obtained from a Coulomb-dissociation experiment~\cite{Moto03} in the range of proton energies 0.2 MeV $\leq  E_{\rm c.m.} \leq$ 0.6 MeV. This systematic feature seen in experimental data is reproduced by the GSM-CC calculation.

In Figs.~\ref{fig-2}-\ref{fig-6}, only the astrophysical factors of the direct capture to the ground state of $^9$C are illustrated. Since the first excited state ${ {1/2}_{1}^{-} }$ of $^9$C is a resonance, its contribution to the total astrophysical $S$ factor will be smaller than that of bound states~\cite{Trache02,BGuo05}.
In Fig.~\ref{fig-7}, the results of the capture to the ground and the first excited state of $^9$C are shown. It is seen that at around $E_{\rm c.m.}$=1 MeV, the explicit contribution of the capture to the ${ {1/2}_{1}^{-} }$ resonance appears. This contribution has a small peak at around  $E_{\rm c.m.}$=1.2 MeV and for higher energies  has a stable contribution of about 20$\%$ to the total astrophysical $S$ factor.

Most of experimental analyses aimed at calculating the astrophysical $S$ factor in the limit $E_{\rm c.m.}\rightarrow$ 0. In Fig. ~\ref{fig-6b}, we depict values of $S(0)$ issued from various experimental analyses and theoretical studies and compare them with the GSM-CC result, denoted as ``Present''. The GSM-CC result shown in this figure is obtained for the standard value of the $E1$ effective charge and the $E2$ effective charges from Ref.~\cite{Hees1988}. The GSM-CC result is compatible with most of the experimental astrophysical $S$ factors. 

\subsection{The astrophysical reaction rate}
\label{sec3-c}
The $^8$B$(p,\gamma)$$^9$C reaction provides an important step in the production of the CNO nuclei in the hot $pp$ chain, especially at temperatures $0.07 \leq {\rm T}_9 \leq 0.7$~\cite{Wiescher89}.
\begin{figure}[htb]
\vskip 0.5truecm
\includegraphics[width=0.9\linewidth]{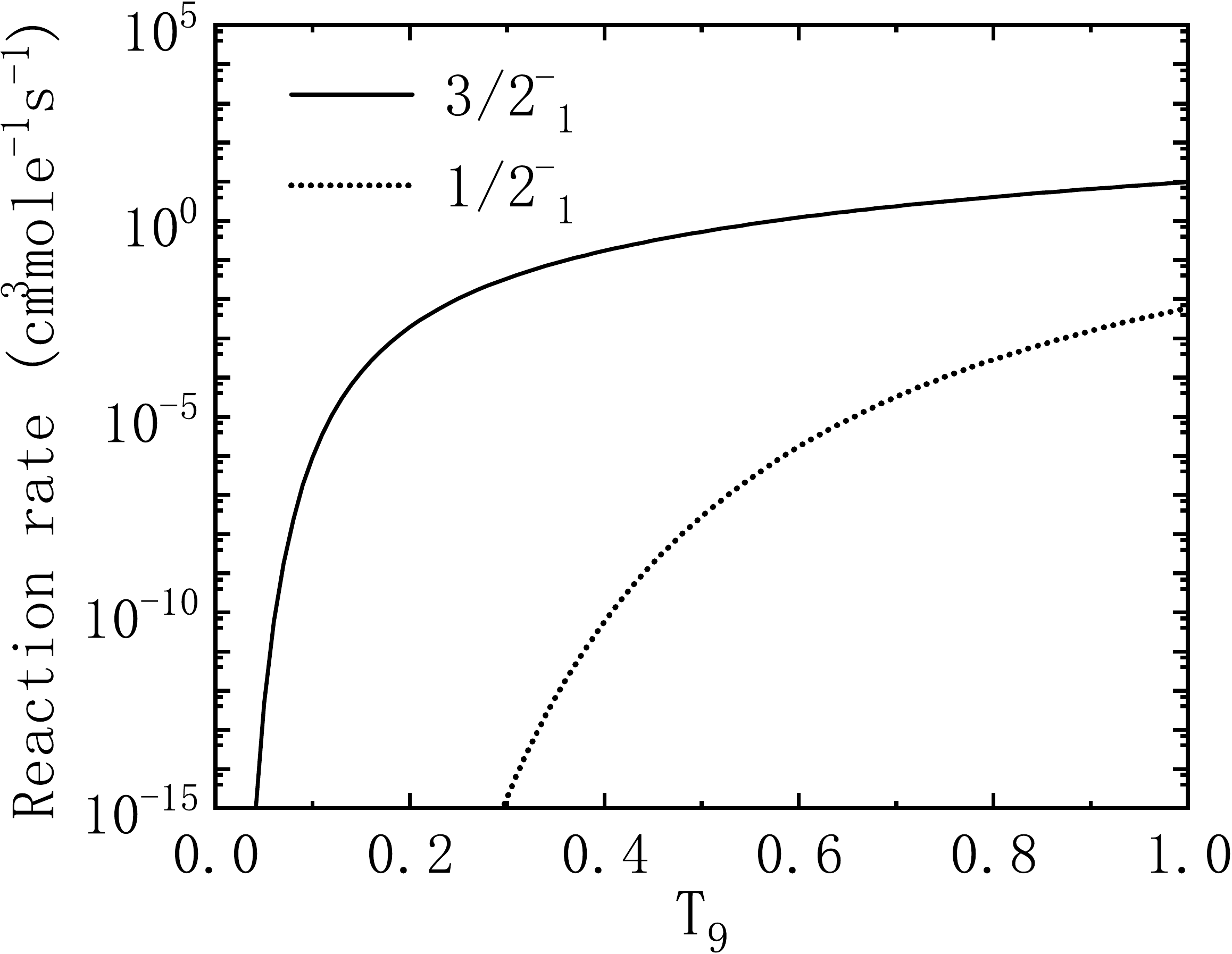}
\caption{The reaction rate of $^8$B$(p,\gamma)$$^9$C as a function of the temperature T$_9$ calculated by GSM-CC. The solid line represents the direct capture rate to the ground state 3/2$^-_1$ and the dotted line depicts the capture to the first excited resonance state 1/2$^-_1$.}
\label{fig-8}
\end{figure}

The astrophysical reaction rates as a function of temperature for the $^8$B$(p,\gamma)$$^9$C reaction calculated by GSM-CC are depicted in Fig.~\ref{fig-8}, for the direct capture to the ground state and the resonant capture to the first excited state of $^9$C. As seen in the figure, the direct capture dominates reaction rates in the range of temperature of astrophysical interest, in agreement with previous studies~\cite{Wiescher89}.

The low reaction rates found in GSM-CC calculations imply that the most efficient temperature for the formation of $^9$C is in the range of $0.14 \leq {\rm T}_9 \leq 0.84 $, i.e.~significantly higher than the estimate given by Wiescher et al.~\cite{Wiescher89} $0.07 \leq {\rm T}_9 \leq 0.7$. 

\section{Conclusions}
\label{sec4}
In low-metallicity supermassive stars, the $^8$B$(p,\gamma)$$^9$C reaction plays an important role in the hot $pp$ chain to produce CNO nuclei~\cite{Wiescher89}. When temperature and density are high, the proton capture of $^8$B can be faster than its beta decay. Due to experimental difficulties and theoretical uncertainties, the reaction rate of $^8$B$(p,\gamma)$$^9$C is not known precisely. The direct measurement of the reaction $^8$B$(p,\gamma)$$^9$C at $E_{\rm c.m.} \sim 0$ is seemingly an impossible task. In indirect measurements, it has been shown that the use of the observed ANC for the calculation of the astrophysical $S$(0) factor is quite uncertain due to strong model dependence~\cite{Trache02,Fukui12,Beaumel01,Fukui15}.

In this work, we have applied the shell model for open quantum systems, the GSM-CC, to investigate both the proton radiative capture cross section of the reaction $^8$B$(p,\gamma)$$^9$C and the temperature dependence of its reaction rate. GSM-CC provides the unified theory of nuclear structure and reactions. In the present work, the Hamiltonian is given by the one-body core potential and the residual two-body interaction whose parameters are adjusted to reproduce the spectra and binding energies of $^8$B and $^{9}$C. Once the Hamiltonian is fixed, the GSM-CC is used to calculate both spectra and cross-sections of the studied reaction using the same many-body approach.

The major contribution to the total astrophysical $S$ factor is given by the $E1$ transitions. Especially, the transition from the proton $s$-wave to the $p$ wave plays the leading role in E1 transitions. In agreement with the cluster model~\cite{Descouvemont93,Descouvemont99}, we find that the p + $^8$B composite continuum states $J_i^{\pi}$= $3/2^+$, $5/2^+$ in $^9$C are most important in $S^{E1}$. Transitions from both these continua show a broad peak at around $E_{\rm c.m.}$ =0.6 MeV. The $M1$ transition contribute less to the astrophysical $S$ factor than the $E1$ transition, but it has a large peak around the energy of ${ {1/2}_{1}^{-} }$ resonance. The contribution from the $E2$ transition to $S$ is much smaller than those of $E1$ and $M1$.

In our calculation, the direct capture to the ground state of $^9$C is dominant in the astrophysical $S$ factor. The capture to the ${ {1/2}_{1}^{-} }$ resonance is less important but provides about a 20$\%$ contribution at higher energies ($E_{\rm c.m.} > $ 1 MeV). The calculation of the temperature dependence of the reaction rate of proton capture shows that, for temperatures of astrophysical interest, direct capture bears the leading contribution.

The existing experimental information about the astrophysical $S$ factor suggest a sharp increase from  $E_{\rm c.m.} \simeq$ 0~\cite{Enders03,Trache02,Beaumel01,BGuo05,Fukui12,Fukui15} to a region 0.2 MeV $\leq E_{\rm c.m.} \leq$ 0.6 MeV, which is probed in Coulomb-dissociation reactions~\cite{Moto03}. This behavior is reproduced in the GSM-CC.

  The GSM-CC value for $S(0)$  is close to the astrophysical factors determined in most indirect measurements~\cite{Enders03,Trache02,Beaumel01,BGuo05,Fukui12,Fukui15}. It is also smaller than most previous theoretical predictions~\cite{Wiescher89,Descouvemont99,Mohr03}. As a consequence, the critical temperature and density at which the proton capture reaction $^8$B$(p,\gamma)$$^9$C becomes faster than beta decay is predicted by GSM-CC to be higher than previously expected.

\section{Acknowledgements}
This work has been supported by the National Natural Science Foundation of China under Grant Nos. U2067205,12275081,12175281, 11605054, and 12147219. G.X.D. and X.B.W. contributed equally to this work.






\end{document}